# POWER-SPECTRUM ANALYSIS OF SUPER-KAMIOKANDE SOLAR NEUTRINO DATA, TAKING INTO ACCOUNT ASYMMETRY IN THE ERROR ESTIMATES

P.A. Sturrock[1]

[1] Center for Space Science and Astrophysics, Varian 302, Stanford University, Stanford, California 94305-4060

## ABSTRACT

Since rotational or similar modulation of the solar neutrino flux would seem to be incompatible with the currently accepted theoretical interpretation of the solar neutrino deficit, it is important to determine whether or not such modulation occurs. There have been several published analyses of the Super-Kamiokande dataset, but these all ignore the asymmetry of the error estimates (the upper error estimate is typically slightly larger than the lower error estimate). The purpose of this article is to carry out a power-spectrum analysis (based on likelihood methods) of the Super-Kamiokande 5-day dataset that takes account of the asymmetry in the error estimates. Whereas the likelihood analysis involves a linear optimization procedure for symmetrical error estimates, it involves a nonlinear optimization procedure for asymmetrical error estimates.

We find that for most frequencies there is little difference between the power spectra derived from analyses of symmetrized error estimates and from asymmetrical error estimates. However, this proves not to be the case for the principal peak in the power spectra, which is found at $9.43\,yr^{-1}$. A likelihood analysis which allows for a "floating offset" and takes account of the start time and end time of each bin and of the flux estimate and the symmetrized error estimate leads to a power of 11.24 for this peak. A Monte Carlo analysis shows that there is a chance of only 1% of finding a peak this big or bigger in the frequency band $1-36\,yr^{-1}$ (the widest band that avoids artificial peaks). On the other hand, an analysis that takes account of the error asymmetry leads to a peak with power 13.24 at that frequency. A Monte Carlo analysis shows that there is a chance of only 0.1% of finding a peak this big or bigger in that frequency band $1-36\,yr^{-1}$. From this perspective, power spectrum analysis that takes account of asymmetry of the error estimates gives evidence for variability that is significant at the 99.9% level.

We comment briefly on an apparent discrepancy between power spectrum analyses of the Super-Kamiokande and SNO solar neutrino experiments.

1. INTRODUCTION

This article is concerned with the analysis of data from the Super-Kamiokande experiment (Fukuda et al., 2001, 2002, 2003) that have been organized in 5-day bins (Yoo et al., 2003). There have been several power-spectrum analyses of Super-Kamiokande data (Yoo et al., 2003; 2003; Milsztajn, 2003; Nakahata et al., 2003; Sturrock, 2003, 2004; Caldwell and Sturrock 2005). To clarify the relationship of these approaches, we have recently presented a sequence of power-spectrum analyses of the Super-Kamiokande 5-day dataset (Sturrock et al., 2005). The most prominent feature in the power spectra is a peak at $9.43\,yr^{-1}$. We found that the strength of this feature increases progressively as we take into account more of the experimental data. One of these analyses gives a power of 11.67 at $9.43\,yr^{-1}$. When we carry out 10,000 Monte Carlo simulations of the data, we find that only 74 simulations have this power or more in the frequency range $0-50\,yr^{-1}$.

The available data for the Super-Kamiokande experiment (Yoo et al., 2003) gives, for each bin, timing data (start time, end time, and mean live time), and the following measurement data: a flux estimate and upper and lower error estimates. The above-mentioned analyses (Yoo et al., 2003; Koshio, 2003; Milsztajn, 2003; Nakahata, 2003; Sturrock, 2003, 2004; Sturrock et al., 2005) have involved the simplifying assumption that the probability distribution function (pdf) for the flux is normal, so that the two error estimates are replaced by a single error estimate. However, the distribution cannot actually be normal, since that would assign non-zero probability to negative values of the flux. This consideration is important for the analysis of radiochemical solar neutrino data for which the error estimates are comparable with the flux estimates (Cleveland et al., 1998; Hampel et al., 1999; Vermul et al., 2002; Altmann et al., 2005), but not so important for the analysis of Super-Kamiokande data, for which the mean of the error estimates is small (20%) in comparison with the mean flux estimate.

In this article, we re-analyze the Super-Kamiokande 5-day dataset, retaining the distinction between the upper and lower error estimates. From the published data (Yoo et al., 2003), we extract the flux $g_r$, which we correct for the varying Sun-Earth distance, and the upper and lower error estimates $\sigma_{ur}$ and $\sigma_{lr}$ (all given in units of $10^6 \, cm^{-2} \, s^{-1}$). The bin number $r$ runs from 1 to $R \, (= 358)$. We show in Figure 1 the pairs of error estimates, normalized with respect to the mean flux ($2.35 \, 10^6 \, cm^{-2} \, s^{-1}$). The mean value of $\sigma_{ur}/mean(g_r)$ is 0.21, and the mean value of $\sigma_{lr}/mean(g_r)$ is 0.18. The ratio $\sigma_{ur}/\sigma_{lr}$ has a minimum value 1.07, a maximum value 1.46, and a mean value 1.182. There is not a big difference between the upper and lower error estimates, so that one would not expect that it will make a huge difference to the resulting power spectrum if, for each bin, we replace the upper and lower error estimates by their mean value (which is the procedure used in prior analyses).

Taking account of the asymmetry is a nontrivial complication, since – as we shall see - it replaces a linear optimization problem with a nonlinear one. We present in Section 2 our analysis of the symmetrized data. We then present in Section 3 an analysis that takes account of the asymmetry of the error estimates. We apply likelihood methods (Sturrock et al., 2005) to both calculations. We discuss the results in Section 4.

## 2. LIKELIHOOD ANALYSIS FOR SYMMETRIZED ERROR ESTIMATES

If $g_r$ is the flux measurement (for each run or bin), normalized to have mean value zero, and if the error distribution is taken to be normal with half-width $\sigma_r$, then the power spectrum may be derived from a likelihood calculation (Sturrock et al., 2005). We first evaluate the likelihood on the assumption that the flux is constant from

$$L_0 = -\tfrac{1}{2} \sum_{r=1}^{R} \frac{(g_r - G_0)^2}{\sigma_r^2} \tag{2.1}$$

where $G_0$ is chosen to maximize $L_0$. (In this and similar expressions, we ignore additive terms that do not involve the parameters (here $G_0$) being estimated.) We then evaluate the likelihood on the assumption that the flux has a sinusoidal modulation from

$$L = -\tfrac{1}{2} \sum_{r=1}^{R} \frac{(g_r - G_r)^2}{\sigma_r^2} \tag{2.2}$$

where $G_r$ is an estimate of the expected flux for sinusoidal modulation, as given by

$$G_r = \frac{1}{t_{er} - t_{sr}} \int_{t_{sr}}^{t_{er}} dt \, W_r(t) \left( K + A e^{i 2\pi \nu t} + A^* e^{-i 2\pi \nu t} \right). \tag{2.3}$$

$W_r(t)$ is a weighting function that takes account of the decay of capture products in radiochemical experiments or live time in Cerenkov experiments. For each frequency, the offset K and the complex amplitude A are adjusted to maximize the likelihood. Then the power S is given by

$$S = L - L_0 . \qquad (2.4)$$

Since S is a quadratic function of K and A, this optimization becomes a simple least-squares problem.

The resulting power sectrum is shown in Figure 2. The top ten peaks are shown in Table 1. We have also generated Monte-Carlo simulations of the data. These simulations are generated as follows: for each bin, we select flux values randomly from a normal probability distribution function centered on the maximum-likelihood $g_{ML}$ estimate of the flux, with width determined by the symmetrized error estimate, and form the power spectrum of this simulation. For each simulation, we then determine the maximum power SM in the frequency range $1 - 36\, yr^{-1}$. We adopt $1\, yr^{-1}$ as the lower limit since the floating offset method breaks down at or near zero frequency (Sturrock et al., 2005). We adopt $36\, yr^{-1}$ as the upper limit to avoid the effects of aliasing caused by the fact that the sampling has a strong periodicity with frequency $72\, yr^{-1}$ (Sturrock, 2004; Sturrock et al., 2005). Figure 3 shows a histogram of SM formed from 10,000 simulations. We find that 98 of 10,000 simulations have values of SM as large as or larger than the actual value, 11.24. From this perspective, the evidence for variability is significant at the 99% level.

## 3. LIKELIHOOD ANALYSIS FOR ASYMMETRIC ERROR ESTIMATES

We now wish to take account of asymmetry of the pdf. We therefore replace (2.1) and (2.2) by

$$L_0 = -\tfrac{1}{2} \sum_{r=1}^{R} \left\{ \frac{(x_r - X_0)^2}{\sigma_{u,r}^2} h(x_r - X_0) + \frac{(x_r - X_0)^2}{\sigma_{l,r}^2} h(-x_r + X_0) \right\} , \qquad (3.1)$$

and

$$L = -\tfrac{1}{2} \sum_{r=1}^{R} \left\{ \frac{(x_r - X_r)^2}{\sigma_{u,r}^2} h(x_r - X_r) + \frac{(x_r - X_r)^2}{\sigma_{l,r}^2} h(-x_r + X_r) \right\} , \qquad (3.2)$$

respectively, where $h(x)$ is the Heaviside function. Since the amplitude A appears implicitly in the Heaviside functions, S is no longer a simple quadratic function of the amplitude, so that the maximization of S becomes a nonlinear problem.

We proceed as follows: We first compute the power S by replacing both upper and lower error estimates by their mean. We may then carry out the power spectrum analysis employed in Section 2. Suppose that the evaluation of $L_0$ by Equation (2.1) leads to an estimate $G_{00}$ of $G_0$. Since the nonlinear calculation gives rise to an estimate of $G_0$ that is not very different from $G_{00}$, we may find the value that maximizes $L_0$ by assuming that $L_0$ is a quadratic function of $G_0$ in the neighborhood of $G_{00}$. Then, by evaluating $L_0$ at $G_{00}$ and at two nearby values, we may find the value of $G_0$ that maximizes $L_0$. We may carry out a similar procedure to find the maximum value of L, adjusting in turn the offset K and the real and imaginary parts of the amplitude A. This procedure could if necessary be iterated to find the values of K and A that minimize L. However, in our analysis of Super-Kamiokande data, we find that iteration is not necessary. The resulting power spectrum is shown in Figure 4. The top ten peaks in the frequency range $1-36\,yr^{-1}$ are listed in Table 2.

By comparing Tables 1 and 2, we see that the maximum power is still found at $9.43\,yr^{-1}$, and that the power has been increased to 13.24. This is in fact the strongest peak in the entire band ($0-100\,yr^{-1}$) for which calculations have been made. The results of 10,000 Monte Carlo simulations, with the search band again chosen to be $1-36\,yr^{-1}$, are shown in histogram form in Figure 5. Only 12 simulations out of 10,000 have peaks as strong as or stronger than the actual power (13.24). From this perspective, power spectrum analysis that takes account of asymmetry of the error estimates gives evidence for variability that is significant at the 99.9% level.

In order to determine whether the change in power at $9.43\,yr^{-1}$ is part of a general pattern or is peculiar to this particular frequency, we show in Figure 6 both power spectra obtained in Sections 2 and 3. We see from this figure that for most frequencies there is very little difference in the results of these two analyses.

## 4. DISCUSSION

We see in Figure 6 that, although there is little difference between the overall power spectra computed with symmetrized error estimates and those computed with asymmetric error estimates, there is a significant difference in the power of the leading peak. In view of the fact that we have obtained this result in an analysis of Super-Kamiokande data, for which there only a small difference between the upper and lower error estimates (see Figure 1), we should probably expect a more pronounced difference in comparable analyses of radiochemical data for which the error asymmetry is more pronounced, so that future analyses of radiochemical datasets should probably take the error asymmetry into account.

A word of caution may be in order concerning the interpretation of power spectra obtained from analyses of solar neutrino datasets. Such analyses are well suited for the detection of a stable, long-lived, high-Q, periodic modulation. Hence there is a temptation to interpret any strong peak in that way. However, the Sun exhibits many periodic modulations, but hardly any are stable, long-lived and high-Q. The solar cycle exhibits a strong periodicity but the period, amplitude and phase vary from cycle to cycle, and there is a well-known interval (the Maunder Minimum) when the solar cycle was not in evidence at all (Eddy, 1976). Most variables that are related to solar radiation or solar activity show a strong periodicity related to solar rotation, but again the modulation will typically vary in amplitude and phase (Castagnoli and Provensal, 1997). The Rieger-type oscillations are particularly erratic. The first of these oscillations was not discovered until 1984 (Rieger et al., 1984). Their characteristics and mechanism are still very much a matter of debate (Ballester, Oliver, and Carbonnel, 2002). A Rieger oscillation may be strong in one cycle and absent in the next (Bai, 2003). Wavelet analyses show that they are typically transient and that the frequency is not constant (Ballester, Oliver, and Carbonnel, 2002; Rybak, Ozguc, Atak, and Sozen, 2005).

These considerations are very relevant to the analysis of solar neutrino data, since the modulations that seem to occur in the neutrino flux are best understood in terms of solar rotation and Rieger-type oscillations, which we propose (Sturrock et al., 1999) may be interpreted as r-mode oscillations (Papaloizou & Pringle, 1978; Provost, Berthomieu, and Rocca, 1981; Saio, 1982). As we have suggested elsewhere (Sturrock, 2004; Caldwell and Sturrock 2005), the peak at $39.28\,yr^{-1}$ may be interpreted as the second harmonic of the solar rotation frequency, and the peaks at $9.43\,yr^{-1}$ and $43.72\,yr^{-1}$ may be attributed to an r-mode oscillation with spherical harmonic indices $l=2, m=2$. For these reasons, one should not expect that the peak at $9.43\,yr^{-1}$ in the power spectrum will, if real, necessarily represent a stable, long-lived, high-Q oscillation.

As a consequence of these considerations, we should view with caution the recent conclusions of the SNO collaboration who fail to find evidence of a modulation at $9.43\,yr^{-1}$ in their data, and conclude that "[their] data are inconsistent with the hypothesis that the results of the recent analysis by Sturrock et al., based on elastic scattering events in Super-Kamiokande, can be attributed to a 7% sinusoidal modulation of the total $^8$Be neutrino flux." (Aharmin et al., 2005). Apart from differences in the measurement processes, there is a significant difference in the intervals examined by the two experiments. The Super-Kamiokande data, analyzed in Sections 2 and 3, span the interval 1996.424 to 2001.535, whereas the SNO D2O and Salt datasets span the intervals 1999.835 to 2001.402 and 2001.566 to 2003.654, respectively.

If we repeat the analysis of Section 2 for the interval that is common to both the Super-Kamiokande and SNO datasets, we find that there is no peak at exactly $9.43\,yr^{-1}$. There is a peak at $9.27\,yr^{-1}$, with power 5.23. By carrying out Monte Carlo simulations, we find a peak with this power or higher in the frequency range $1-36\,yr^{-1}$ for 452 trials out of 1000. Hence the fact that SNO data do not reveal a peak at $9.43\,yr^{-1}$ is quite compatible with what we find in the Super-Kamiokande power spectrum for the same time interval. The absence of a peak at $9.43\,yr^{-1}$ in the SNO power spectrum has little bearing on the significance of the peak at that frequency in the complete Super-Kamiokande dataset. It would probably be more appropriate, and may avoid future confusion, if one were to use the term "quasi-periodic" in referring to modulations that appear in analyses of solar neutrino data. One would obtain a better understanding of these modulations by using time-frequency analysis, which we plan to do in the near future. We plan also to carry out a more detailed comparison of SNO and Super-Kamiokande data.

This work was supported by NSF grant AST-0097128. I wish to thank the Super-Kamiokande and SNO consortia for making their data publicly available, and D. O. Caldwell, J. Pulido, and J. Scargle for helpful discussions.

## TABLE 1

Top ten peaks in a likelihood power spectrum computed from the 5-day dataset, for the frequency range $1-50\,yr^{-1}$, calculated using start times and end times, adopting a symmetrized error estimates, and allowing for a floating offset.

| Order | Frequency (yr$^{-1}$) | Power |
|---|---|---|
| 1 | 9.43 | 11.24 |
| 2 | 43.72 | 9.44 |
| 3 | 39.28 | 8.64 |
| 4 | 48.43 | 6.38 |
| 5 | 45.86 | 6.10 |
| 6 | 31.24 | 6.03 |
| 7 | 12.31 | 6.01 |
| 8 | 48.16 | 5.69 |
| 9 | 33.99 | 5.63 |
| 10 | 39.55 | 5.32 |

## TABLE 2

Top ten peaks in a likelihood power spectrum computed from the 5-day dataset, for the frequency range $1-50\,yr^{-1}$, calculated using start times and end times, taking account of the asymmetric error estimates, and allowing for a floating offset.

| Order | Frequency (yr$^{-1}$) | Power |
|---|---|---|
| 1 | 9.43 | 13.24 |
| 2 | 43.72 | 9.97 |
| 3 | 39.28 | 9.00 |
| 4 | 48.43 | 7.42 |
| 5 | 45.85 | 7.18 |
| 6 | 48.16 | 6.35 |
| 7 | 12.31 | 6.24 |
| 8 | 37.12 | 6.10 |
| 9 | 39.54 | 5.67 |
| 10 | 8.29 | 5.60 |

FIGURES

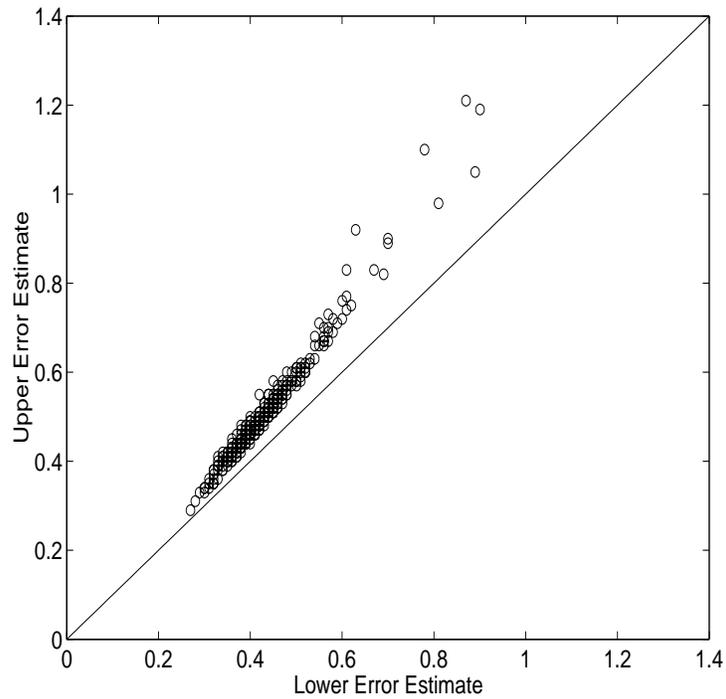

Figure 1. Upper and lower error estimates for the Super-Kamiokande 5-day dataset.

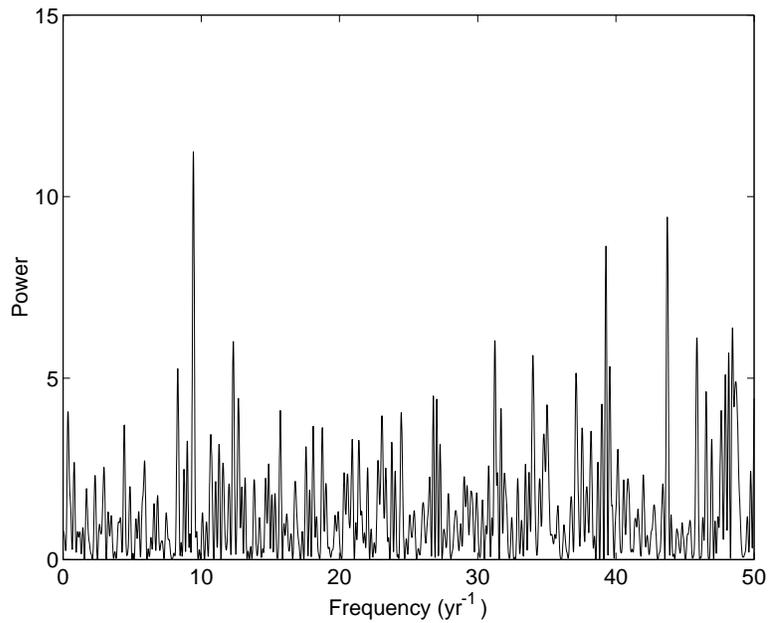

Figure 2. Power spectrum computed by a "floating-offset" likelihood analysis that incorporates symmetrized error estimates.

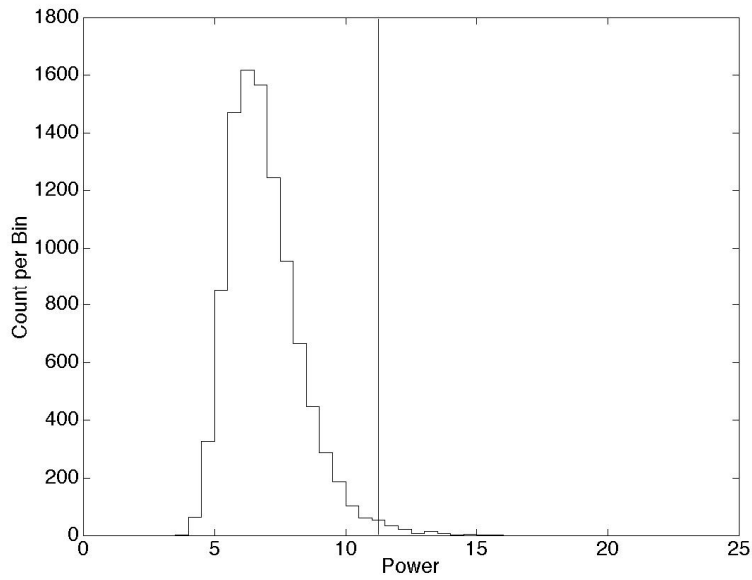

Figure 3. Histogram display of the maximum power, computed by the likelihood method using the start times and end times and allowing for a floating offset, over the frequency band 1 to 36 $yr^{-1}$, for 10,000 Monte Carlo simulations of the Super-Kamiokande 5-day data. 98 out of 10,000 simulations have power larger than the actual maximum power (11.24 at frequency 9.43 $yr^{-1}$).

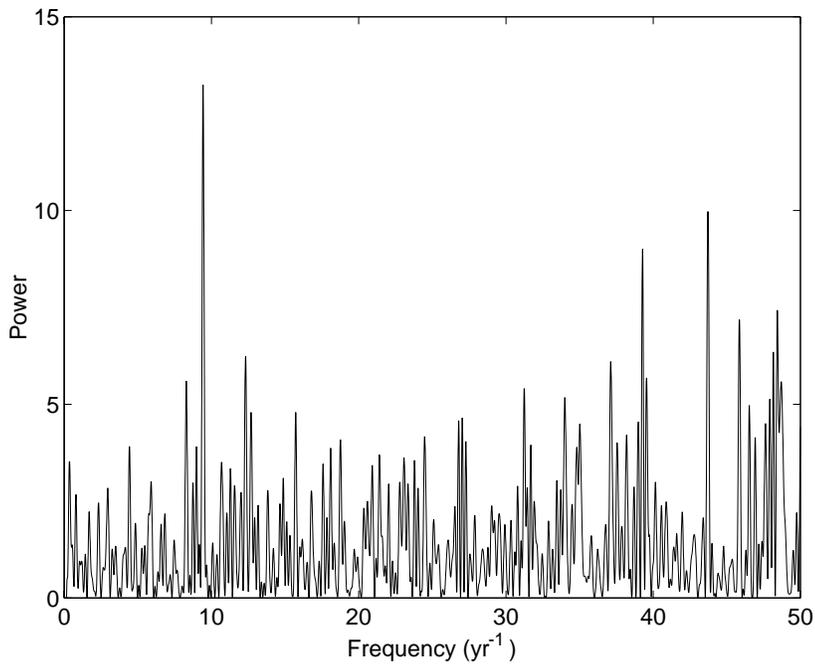

Figure 4. Power spectrum computed by a likelihood procedure that takes account of the asymmetry in the error estimates.

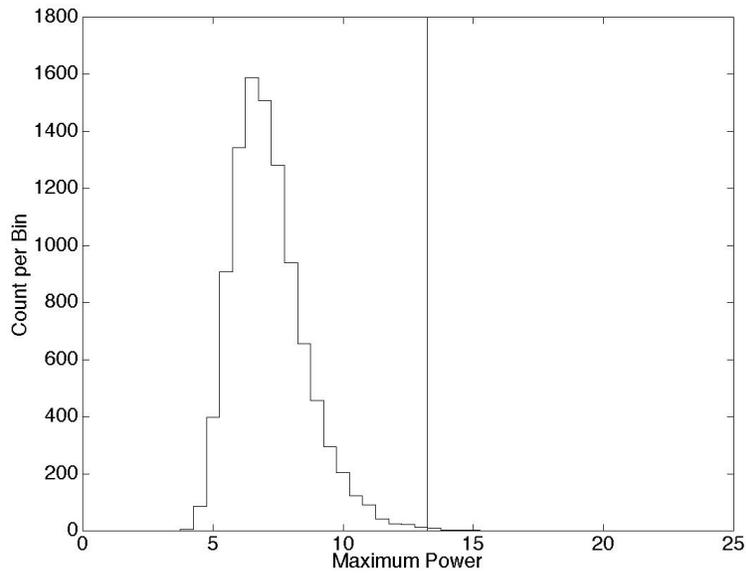

Figure 5. Histogram display of the maximum power, over the frequency band 1 to 36 yr$^{-1}$, computed by the likelihood method using the start times and end times, allowing for a floating offset, and taking account of the asymmetry in the error estimates, for 10,000 Monte Carlo simulations of the Super-Kamiokande 5-day data. Only 11 simulations out of 10,000 have power larger than the actual maximum power (13.15 at frequency 9.43 yr$^{-1}$).

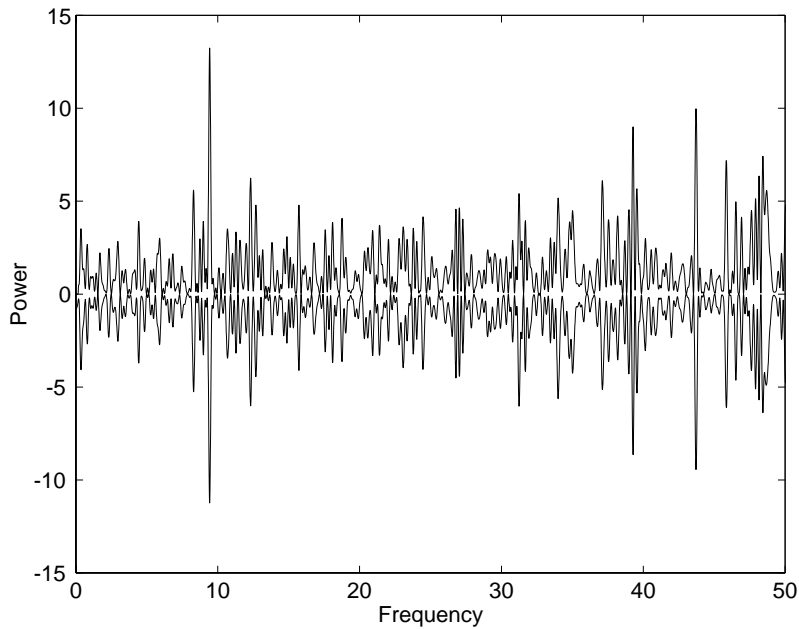

Figure 6. Comparison of power spectra computed from asymmetric error estimates (shown positive) and from symmetrized error estimates (shown negative).